\documentclass[prd,reprint,showpacs,showkeys]{revtex4-1}
\usepackage{graphics}
\usepackage{eurosym}
\usepackage{amsfonts}
\usepackage{amssymb}
\usepackage{amsmath}
\usepackage{graphicx}
\usepackage[font={footnotesize,it}]{caption}
\usepackage{hyperref}

\begin{document}

\title{Quantum Tunneling of Spin-1 Particles from a 5D Einstein-Yang-Mills-Gauss-Bonnet Black Hole Beyond Semiclassical Approximation}
\author{Kimet Jusufi}
\email{kimet.jusufi@unite.edu.mk}
\affiliation{Physics Department, State University of Tetovo, Ilinden Street nn, 1200, Tetovo,
Macedonia}
\date{\today }

\begin{abstract}
In the present paper we study the Hawking radiation as a quantum tunneling effect of spin--$1$ particles from  a five-dimensional, spherically symmetric, Einstein--Yang--Mills--Gauss-Bonnet (5D EYMGB) black hole. We solve the Proca equation (PE) by applying the WKB approximation and separation of variables via Hamilton--Jacobi (HJ) equation which results in a set of five differential equations, and reproduces in this way, the Hawking temperature.  In the second part of this paper, we extend our results beyond the semiclassical approximation. In particular, we derive the logarithmic correction to the entropy of the 5D EYMGB black hole and show that the quantum corrected specific heat indicates the possible existence of a remnant.
\end{abstract}
\maketitle

\section{Introduction}

Hawking radiation \cite{hawking1,hawking2,hawking3}, can be investigated in many different ways. One such an interesting method, is the so-called tunneling method, also known as a Kraus--Parkih--Wilczek method \cite{kraus1,kraus2,perkih1,perkih2,perkih3}. Recently, the Hamilton--Jacobi (HJ) method of tunneling was extended  by Padmanabhan et al \cite{sri1,sri2}. There are some interesting aspects associated about this method, for example, the whole process is consistent with an underlying unitary theory due to the conservation of energy  which may have implications in the information loss paradox. This method has been extensively studied in the literature \cite{mann0,mann1,mann2,mann3,gohar1,ahmed1}. Note that Hawking radiation in the tunneling approach, emerges in the regime of semi classical approximation, thus naturally, Banerjee and Majhi \cite{banerjee1,banerjee2,akbar,corda}, extended this method beyond the semiclassical approximation by including higher order quantum corrections to the Hawking temperature.  While, many other authors recently studied the role of quantum gravity effects on the Hawking radiation by exploring the deformed Klein--Gordon equation, deformed Dirac equation, deformed Wheeler--DeWitt equation and shown that quantum gravity effects can decelerate the evaporation process of the black hole and as a result remnants are left \cite{qgv1,qgv2,brito1,brito2,khalil1,khalil2,khalil3,khalil4,khalil5}.

Besides the well known solutions of Einstein field equations, such as, a vacuum solution, charged black hole solution, spinning black hole, it came as a surprise when other solutions were shown to exist in the realm of Yang-Mills (YM) theory coupled with gravity known as Einstein--Yang--Mills theory (EYM) \cite{betti,maeda,chak,wu}. In Ref. \cite{mazh1,mazh2} EYM black holes in Gauss--Bonnet theory was studied, in Ref. \cite{kunz} EYM theory with adjoint Higgs field was investigated, in Ref. \cite{yves} EYM black hole in Chern--Simons theory was studied. While the Hawking radiation and other thermodynamic aspects of EYM black holes were investigated in Ref \cite{okyama,gosh,mazh3,Konoplya,odintsov1,odintsov2,chak1,chak2,martin,safia}

The Hawking radiation of massive gauge bosons, such as $W^{\pm}$ and $Z$ bosons which carry the weak interaction was analysed in Ref. \cite{kruglov1,xiang,sakalli1,sakalli2,sakalli3,kimet}. Recently, massive vector particles  have attracted interest, in particular massive photons were proposed as a possible explanation of the dark energy and dark matter \cite{seyen, kahn,bednyakov}, and more recently the so-called ultra--light bosons were introduced as a possible explanation of the dark matter \cite{ed}. There is a common belief that the Hawking temperature is usually determined by the black hole mass, charge, and the angular momentum of the black hole and this temperature is unaffected by the mass of the particles. 

In this paper we aim to extend the Hawking radiation of massive and uncharged vector particles as a quantum tunneling effect to the 5D EYMGB black hole solution \cite{gosh} and to determine whether the mass and the spin of the particles plays any relevant role in the Hawking temperature in the case of 5D EYMGB black holes. In particular we also aim to incorporate the quantum effects beyond the semiclassical approximatin on the Hawking temperature, entropy and specific heat capacity from this black hole configuration. As far as we know, this problem has not yet been addressed in the literature. 
The paper is organized as follows. In Section II, we review the 5D EYM solution in Gauss--Bonnet gravity. In Section III, we solve the Proca equatition and recover the Hawking temperature from the 5D EYMG black hole. In Section IV, we extend our results beyond the semiclassical approximation by computing the corrected entropy and specific heat capacity of the black hole. In Section V, we comment on our results.
\bigskip
\section{5D Spherically Symmetric EYMGB Black Hole Solution }

Let us start by writing the action which describes the  5D EMYGB theory given by \cite{mazh1,mazh2,gosh}
\begin{equation}
S=\frac{1}{2}\int_{\mathcal{M}}\sqrt{-g}\left(R+\alpha \mathcal{L}_{GB}-\sum_{a=1}^{6}F_{\mu\nu}^{(a)}F^{(a)\mu\nu} \right)\mathrm{d}^{5}x,\label{1}
\end{equation}
in which $g = \det(g_{\mu\nu})$ is the determinant of the metric tensor, $\alpha$ is  Gauss-Bonnet (GB) parameter. The GB Lagrangian $\mathcal{L}_{GB}$, is given by
\begin{equation}
\mathcal{L}_{GB}=R^{2}-4R_{\alpha \beta} R^{\alpha \beta}+R_{\mu\nu\alpha\beta}R^{\mu\nu\alpha\beta}.\label{2}
\end{equation}

Variation of the action (\ref{1}) with respect to the metric $g_{\mu\nu}$ leads to the following modified Einsteins field equations given by \cite{mazh2}
\begin{equation}
G_{\mu\nu}-\alpha H_{\mu\nu}=T_{\mu\nu},\label{3}
\end{equation}
where $G_{\mu\nu}$ is the Einstein tensor, $H_{\mu\nu}$ is a symmetric rank--two tensor related to the spacetime geometry usually called Lovelock tensor. The energy--momentum tensor $T_{\mu\nu}$, for the Yang--Mills field is given by
\begin{equation}
T_{\mu\nu}=2{{F^{i}}_{\mu}}^{\alpha}{F^{i}}_{\nu\alpha}-\frac{1}{2}g_{\mu\nu}{F^{i}}_{\alpha\beta}F^{i\alpha\beta},\label{4}
\end{equation}
in which ${F^{i}}_{\alpha\beta}F^{i\alpha\beta}=\frac{6Q^{2}}{r^{4}}$ and $Q$ is the only non--zero gauge charge and $F^{i}_{\alpha\beta}$ is the Yang-Mills field 2--forms. The Lovelock tensor $H_{\mu\nu}$ is defined as \cite{mazh2,chak}
\begin{eqnarray}\nonumber
H_{\mu\nu}&=&\frac{1}{2}g_{\mu\nu}\left(R_{\alpha \beta \gamma \delta}R^{\alpha \beta \gamma \delta}-4R_{\alpha \beta}R^{\alpha \beta}+R^{2}\right)-2RR_{\mu\nu}\\
&+&{4R_{\mu}}^{\lambda}R_{\lambda \nu}+4R^{\rho \sigma}R_{\mu \rho \nu \sigma}-{R_{\mu}}^{\alpha \beta \gamma}R_{\nu \alpha \beta \gamma}.\label{5}
\end{eqnarray}

The 5D solution for spherically symmetric spacetime in EYMGB theory has been derived by Mazharimousavi and Halilsoy \cite{mazh2}
\begin{equation}
\mathrm{d}s^{2}=-F(r)\mathrm{d}t^{2}+\frac{\mathrm{d}r^{2}}{F(r)}+r^{2}\mathrm{d}\Omega_{3}^{2},\label{6}
\end{equation}
where the unit three sphere $\mathrm{d}\Omega_{3}^{2}$ can be written in terms of Euler angles as (see, for example Eq. (4) and Eq. (5) in Ref. \cite{mazh2})
\begin{equation}
\mathrm{d}\Omega_{3}^{2}=\frac{1}{4}\left(\mathrm{d}\theta^{2}+\mathrm{d}\phi^{2}+\mathrm{d}\psi^{2}-2 \cos\theta \,\mathrm{d}\phi \mathrm{d}\psi \right), \label{7}
\end{equation}
where $0 \leq\theta \leq \pi, \,\,\,0\leq\phi,\psi \leq 2\pi $. Morover the above non--vanishing components of field equations is shown to have the following solution \cite{mazh2}
\begin{equation}
F(r)=1+\frac{r^{2}}{4 \alpha}\pm \sqrt{\left(\frac{r^{2}}{4\alpha}\right)^{2}+\left(1+\frac{M}{2 \alpha}\right)+\frac{Q^{2}\ln r}{\alpha}},\label{8}
\end{equation}
in which $M$ is the usual integration constant to be identified as mass. Choosing the minus sign in the limit $\alpha\to 0$, it is possible to recover the EYM solution (see, for example \cite{okyama})
\begin{equation}
F(r)=1-\frac{M}{r^{2}}-\frac{2Q^{2}}{r^{2}}\ln r. \label{9}
\end{equation}

The radius of the event horizon $r_{h}$ can be found by considering  $F(r_{h})=0$, which can be reduced to the following simple equation
\begin{equation}
r^{2}_{h}-M-2Q^{2}\ln(r_{h})=0. \label{10}
\end{equation}

One can observe that the event horizon remains unaffected by $\alpha$ parameter. The corresponding black hole temperature be calculated as 
\begin{equation}
T_{H}=\frac{\kappa}{2\pi}
\end{equation}
in which the surface gravity $\kappa$ is given by \cite{mazh2}
\begin{equation}
\kappa=\left|\frac{1}{2}F^{\prime}(r_{h})\right|.
\end{equation}

The result for the Hawking temperature can be simplified if we set $M=1$ and $Q<1$, which implies that the event horizon is $r_{h}=1$. Furthermore in the units $G=c=1$, the Hawking temperature can be calculated as \cite{mazh2}
\begin{equation}
T_{H}=\frac{\hbar}{2\pi}\frac{1-Q^{2}}{4\alpha+1}. \label{11}
\end{equation}

In the next section we are going to use the metric (\ref{6}) to study the Hawking radiation of massive bosons via quantum tunneling.

\section{Tunneling of massive spin-1 particles from 5D Spherically Symmetric EYMGB Black Hole}

The equation of motion which describes a massive and uncharged boson field $\Psi^{\mu}$, is given by Proca
equation (PE) as follows \cite{kruglov1}
\begin{eqnarray}
\frac{1}{\sqrt{-g}}\partial_{\mu}\left(\sqrt{-g}\,\Psi^{\mu\nu}\right)-\frac{m^{2}}{\hbar^{2}}\Psi^{\nu}=0,\label{12}
\end{eqnarray}
where 
\begin{equation}
\Psi_{\mu\nu}=\partial_{\mu}\Psi_{\nu}-\partial_{\nu}\Psi_{\mu}.\label{13}
\end{equation}

We can now solve PE equation in the spacetime given by the line element (\ref{5}). Therefore, let us start by applying the WKB approximation method which suggests that
\begin{eqnarray}\nonumber
\Psi_{\nu}&=&C_{\nu}(t,r,\theta,\phi,\psi)\exp\Big(\frac{i}{\hbar}\big(S_{0}(t,r,\theta,\phi,\psi)\\
&+&\hbar \, S_{1}(t,r,\theta,\phi,\psi)+\dots \big)\Big).\label{14}
\end{eqnarray}

Furthermore by considering the spacetime symmetries of the metric (\ref{6}), the following ansatz for the action can be choosen
\begin{equation}
S_{0}(t,r,\theta,\phi,\psi)=-Et+R(r,\theta)+j\phi+l\psi,\label{15}
\end{equation}
in which $E$ is the energy of the particle, $j$ and $l$ denotes the angular momentum of the particle corresponding to the angles $\phi$ and $\psi$, respectively. If we now insert the Eq. (\ref{14}) into the Eq. (\ref{12}) and keep only the leading order terms in $\hbar$ we find the following set of five differential equations:
\bigskip
\begin{eqnarray}\nonumber\label{16}
0&=&E R(r)^{\prime}C_{1}+\frac{4 E (\partial_{\theta}R)C_{2}}{r^{2}F(r)}+\frac{4E(\cos \theta l+j)C_{3}}{r^{2}F(r)\sin^{2}\theta}\\
&&+\frac{4E(\cos \theta j+l)C_{4}}{r^{2}F(r)\sin^{2}\theta}+\frac{C_{5}}{r^{2}F(r)\sin^{2}\theta}\mathcal{P}_{1},\end{eqnarray}
\bigskip
\begin{eqnarray}
0&=&\frac{C_{1}}{r^{2}\sin^{2}\theta}\mathcal{P}_{2}+\frac{4R^{\prime}(\partial_{\theta}R)F(r)C_{2}}{r^{2}}+E R^{\prime}C_{5}\\\nonumber
&+&\frac{4 R^{\prime}F(r)}{r^{2}\sin^{2}\theta}\left(\cos\theta l+j\right)C_{3}+\frac{4 R^{\prime}F(r)}{r^{2}\sin^{2}\theta}\left(\cos\theta j+l\right)C_{4},\label{17}
\end{eqnarray}
\begin{eqnarray}\nonumber
0&=& \frac{4 F(\partial_{\theta}R) R^{\prime}}{r^{2}} C_{1}+\frac{4 C_{2}}{r^{4}F(r)\sin^{2}\theta}\mathcal{P}_{3}\\\nonumber
&&+\frac{16 (\partial_{\theta}R)}{r^{4}\sin^{2}\theta}\left(\cos\theta l+j\right)C_{3}+\frac{16 (\partial_{\theta}R)}{r^{4}\sin^{2}\theta}\left(\cos\theta j+l\right)C_{4}\\
&&+\frac{4 E (\partial_{\theta}R)C_{5}}{r^{2}F(r)},\label{18}
\end{eqnarray}
\begin{eqnarray}\nonumber
0&=& \frac{4 F(r)R^{\prime}(r)}{r^{2}\sin^{2}\theta}\left(\cos\theta l+j\right)C_{1}+\frac{16 (\partial_{\theta}R)}{r^{4}\sin^{2}\theta}\left(\cos\theta l+j\right)C_{2}\\\nonumber
&&+\frac{4 E}{r^{2} F \sin^{2}\theta}\left(\cos\theta l+j \right)C_{5}-\frac{4 C_{4}}{r^{4}F\sin^{2}\theta}\mathcal{P}_{4}\\
&& -\frac{4 C_{3}}{r^{4}F \sin^{2}\theta}\mathcal{P}_{5},\label{19}
\end{eqnarray}
\begin{eqnarray}\nonumber
0&=&\frac{4 F R^{\prime}}{r^{2} \sin^{2}\theta}\left(\cos\theta j+l\right)C_{1}-\frac{4 C_{3}}{r^{4}F\sin^{2}\theta}\mathcal{P}_{6}\\\nonumber
&&+\frac{16 (\partial_{\theta}R)}{r^{4}\sin^{2}\theta}\left(\cos\theta j+l\right)C_{2}-\frac{4C_{4}}{r^{4}\sin^{2}\theta F}\mathcal{P}_{7}\\&&+\frac{4E}{r^{2}F \sin^{2}\theta}\left(\cos\theta j+l\right)C_{5}.\label{20}
\end{eqnarray}
where 
\begin{eqnarray}\nonumber
\mathcal{P}_{1}&=&F(r)r^{2}(R^{\prime})^{2}\sin^{2}\theta+m^{2}r^{2}\sin^{2}\theta+8\cos\theta j l \\\nonumber
&&+4(\partial_{\theta}R)^{2}\sin^{2}\theta+4j^{2}+4l^{2},
\end{eqnarray}
\begin{eqnarray}\nonumber
\mathcal{P}_{2}&=&E^{2}r^{2}\sin^{2}\theta-F(r)m^{2}r^{2}\sin^{2}\theta-4(\partial_{\theta}R)^{2}F(r)\sin^{2}\theta\\\nonumber
&-&8\cos\theta F(r) j l-4F(r)j^{2}-4F(r)l^{2},
\end{eqnarray}
\begin{eqnarray}\nonumber
\mathcal{P}_{3}&=&E^{2}r^{2}\sin^{2}\theta-(R^{\prime})^{2}F^{2}r^{2}\sin^{2}\theta-F m^{2}r^{2}\sin^{2}\theta \\
&&-8 \cos\theta Fj l-4Fj^{2}-4Fl^{2},\nonumber
\end{eqnarray}
\begin{eqnarray}\nonumber
\mathcal{P}_{4}&=&\mathcal{P}_{6}=\cos\theta (R^{\prime})^{2}F^{2}r^{2}+\cos\theta Fm^{2}r^{2}-E^{2}\cos\theta r^{2}\\
&&+4 \cos\theta (\partial_{\theta}R)^{2}F-4F jl\nonumber,
\end{eqnarray}
\begin{eqnarray}\nonumber
\mathcal{P}_{5}=(R^{\prime})^{2}F^{2}r^{2}+Fm^{2}r^{2}-E^{2}r^{2}+4(\partial_{\theta}R)^{2}F+4Fl^{2}
\end{eqnarray}
\begin{eqnarray}\nonumber
\mathcal{P}_{7}=(R^{\prime})^{2}Fr^{2}+F m^{2}r^{2}-E^{2}r^{2}+4(\partial_{\theta}R)^{2}F+4Fj^{2}.
\end{eqnarray}

From Eqs. (\ref{16}--\ref{20}) it's clear that one can construct a matrix equation if we introduce a $5\times 5$ matrix, say $\Xi$, which when multiplied by the transpose of a vector $(C_{1}, C_{2}, C_{3}, C_{4}, C_{5})$ gives the following matrix equation

\begin{equation}
\Xi(C_{1}, C_{2}, C_{3}, C_{4}, C_{5})^{T}=0.
\end{equation}

The non--zero elements of the matrix $\Xi$ are given by: 
\bigskip
\begin{eqnarray}\nonumber
\Xi_{11}&=&\Xi_{25}=E R^{\prime},\\\nonumber
\Xi_{12}&=&\Xi_{35}=\frac{4 E (\partial_{\theta}R)}{r^{2}F(r)},\\\nonumber
\Xi_{13}&=&\Xi_{45}=\frac{4E(\cos \theta l+j)}{r^{2}F(r)\sin^{2}\theta},\\\nonumber
\Xi_{14}&=&\Xi_{55}=\frac{4E(\cos \theta j+l)}{r^{2}F(r)\sin^{2}\theta},\\\nonumber
\Xi_{15}&=& \frac{1}{r^{2}F(r)\sin^{2}\theta}\mathcal{P}_{1},\\\nonumber
\Xi_{21}&=&\frac{1}{r^{2}\sin^{2}\theta}\mathcal{P}_{2},\\\nonumber
\Xi_{22}&=&\Xi_{31}= \frac{4R^{\prime}(\partial_{\theta}R)F(r)}{r^{2}},\\\nonumber
\Xi_{23}&=&\Xi_{41}=\frac{4 R^{\prime}F(r)}{r^{2}\sin^{2}\theta}\left(\cos\theta l+j\right),\\\nonumber
\Xi_{24}&=&\Xi_{51}=\frac{4 R^{\prime}F(r)}{r^{2}\sin^{2}\theta}\left(\cos\theta j+l\right),\\\nonumber
\Xi_{32}&=& \frac{4}{r^{4}F(r)\sin^{2}\theta}\mathcal{P}_{3},\\\nonumber
\Xi_{33}&=&\Xi_{42}= \frac{16 (\partial_{\theta}R)}{r^{4}\sin^{2}\theta}\left(\cos\theta l+j\right),\\\nonumber
\Xi_{34}&=&\Xi_{52}=\frac{16 (\partial_{\theta}R)}{r^{4}\sin^{2}\theta}\left(\cos\theta j+l\right),\\\nonumber
\Xi_{43}&=&-\frac{4 }{r^{4}F \sin^{2}\theta}\mathcal{P}_{5},\\\nonumber
\Xi_{44}&=&\Xi_{53}= -\frac{4}{r^{4}F\sin^{2}\theta}\mathcal{P}_{6},\nonumber\\
\Xi_{54}&=&-\frac{4}{r^{4}F\sin^{2}\theta }\mathcal{P}_{7}.
\end{eqnarray}
\bigskip

Since we are looking for the radial trajectories  $R^{\prime}(r)$ of the emitted particle from the black hole, we can find this differential equation from the condition $\det\Xi = 0$, which  leads to the following simplifed equation: 
\begin{equation}
\frac{64 m^{2} \Big(\Theta-E^{2}r^{2}\sin^{2}\theta+F(r)\Omega \Big)^{4}}{r^{14}F^{4}(r)\sin^{10}\theta}=0,
\end{equation}
where
\begin{eqnarray}\nonumber
\Theta=4\sin^{2}\theta F(r)(\partial_{\theta}R)^{2}+F^{2}(r)r^{2}\sin^{2}\theta (R^{\prime})^{2}
\end{eqnarray}
and 
\begin{eqnarray}\nonumber
\Omega=\left(m^{2}r^{2}\sin^{2}\theta+4j^{2}+4l^{2}+8jl\cos\theta\right).
\end{eqnarray}

On can easily solve the last equation for the radial part to get the following integral 
\begin{equation}\label{26}
R_{\pm}(r)=\pm \int \frac{\sqrt{E^{2}-F(r)\left( m^{2}+4 \Sigma \right)}}{F(r)}\mathrm{d}r.
\end{equation}
where
\begin{equation}
\Sigma=(\partial_{\theta}R)^{2}+\frac{1}{r^{2}\sin^{2}\theta}(j^{2}+l^{2}+2 j l\cos\theta).
\end{equation}

A careful analyzes shows that, there is an ambiguity of a factor two associated with the solution of the above problem. This factor of two problem with the original tunneling method and the issue of canonical invariance under canonical transformations given by  $\oint p_{r}\mathrm{d}r=\int p_{r}^{+}\mathrm{d}r-\int p_{r}^{-}\mathrm{d}r$, where $p_{r}^{\pm}=\pm\partial_{r}R$,  was pointed out first in Ref. \cite{Akhmedova0,Akhmedova1,borun}. The resolution of this problem is that there is a temporal contribution to the imaginary part of the tunneling amplitude which was first pointed out in Ref. \cite{Akhmedova2,Akhmedova3,Akhmedova4}. In order to solve this integral, we see that there is a pole at the horizon $r = r_{h}$, since $F(r_{h}) = 0$. We can shift the pole into the upper / lower  half plane $r_{h}\to r_{h}\pm i\epsilon$ and using a Taylor expansion of $F(r)$ near the horizon then the integral (\ref{26}) reads
\begin{equation}\label{int}
R_{\pm}(r)=\pm \lim_{\epsilon \to 0} \int \frac{\sqrt{E^{2}-F(r)\left( m^{2}+4 \Sigma \right)}}{F^{\prime}(r_{h}){(r-r_{h}\pm i\epsilon)}}\mathrm{d}r.
\end{equation}

If we now make use of the equation
\begin{equation}
\lim_{\epsilon \to 0}\text{Im}\frac{1}{r-r_{h}\pm i\epsilon}=\pi \delta (r-r_{h}),
\end{equation}
then Eq. (\ref{int}) gives
\begin{equation}
\text{Im} R_{\pm}(r)=\pm \frac{\pi E}{F^{\prime}(r_{h})},
\end{equation}
where
\begin{equation}
F^{\prime}(r_{h})=\left|\frac{r_{h}}{2\alpha}-\left(\frac{r_{h}^{3}}{4\alpha^{2}}+\frac{Q^{2}}{r_{h}\alpha}\right)\frac{1}{\sqrt{\zeta}}\right|,
\end{equation}
and
\begin{equation}
\zeta=\frac{r_{h}^{4}}{4\alpha^{2}}+4+\frac{2M}{\alpha}+\frac{4 Q^{2}\ln(r_{h})}{\alpha}.
\end{equation}

The spatial contribution to the tunneling can be calculated as
\begin{eqnarray}\nonumber
\Gamma_{spatial}&\propto &\exp\left(-\frac{1}{\hbar}\text{Im} \oint p_{r} \mathrm{d}r\right)\\\nonumber
&=&\exp\left[-\frac{1}{\hbar} \text{Im} \left(\int p_{r}^{+}\mathrm{d}r-\int p_{r}^{-}\mathrm{d}r\right) \right]\\
&=& \exp \left(-\frac{2 \pi E}{\hbar F^{\prime}(r_{h})}\right).
\end{eqnarray}

The temporal part contribution on the other hand, comes due to the connection of the interior region and the exterior region of the black hole. If one introduces $t\to t -i\pi/(2\kappa) $ we will have Im($E\Delta t^{out,in})=-E\pi/(2\kappa)$. Then the total temporal contribution for a round trip gives
\begin{eqnarray}\nonumber
\Gamma_{temporal}&\propto &\exp\left[\frac{1}{\hbar}\left(\text{Im}(E\Delta t^{out})+\text{Im}(E\Delta t^{in}\right)\right]\\
&=&\exp \left(-\frac{2 \pi E}{\hbar F^{\prime}(r_{h})}\right),
\end{eqnarray}
since 
\begin{equation}
\kappa =\left|\frac{1}{2}F^{\prime}(r_{h})\right|.
\end{equation}

One can now define the total tunneling rate of the particles tunneling from inside to outside the horizon  as follows
\begin{eqnarray}\nonumber
\Gamma &=& \exp \Big[\frac{1}{\hbar}\Big(\text{Im}(E\Delta t^{out})+\text{Im}(E\Delta t^{in})\\
&&-\text{Im}\oint p_{r} dr \Big)\Big]= \exp \left[- \frac{4 \pi E}{\hbar F^{\prime}(r_{h})}\right].
\end{eqnarray}

We can now follow \cite{mazh2}, by setting $M=1$ and $Q<1$ as in Section I and by compering the last equation with the Boltzmann factor $\Gamma_{B}=e^{-\beta E}$, after some approximations we find the following result
\begin{equation}
T_{H}=\frac{\hbar}{2 \pi}\frac{1-Q^{2}}{4\alpha+1}.
\end{equation}

The last expression for the Hawking temperature is similar to Eq. (\ref{11}). In other words, we have recovered the Hawking temperature using the quantum tunneling method for 5D EYMGB black hole measured by some observer at infinity.

\section{Temperature and Entropy correction Beyond Semiclassical Approximation}

We can now move on and take into account the quantum effects to the Hawking temperature, entropy, and specific heat capacity of 5D EYMGB black hole. We recall that by following Banerjee and Majhi \cite{banerjee1,banerjee2,akbar,corda}, one may write the action of the particle in the following way 
\begin{equation}\label{action}
S(t,r)=S_{0}(t,r)+\sum_{i}\hbar^{i}S_{i}. \,\,\,\, \left(i=1,2,3,\dots \right).
\end{equation}

It was shown by Banerjee and Majhi that $S_{i}$ are proportional to $S_{0}$, i.e. $S_{i}\propto S_{0}$ \cite{banerjee1,banerjee2}. Therefore from the last equation it follows that
\begin{eqnarray}
S(t,r)=\left(1+\sum_{i}\gamma_{i}\hbar^{i}\right) S_{0}(r,t).
\end{eqnarray}

Since $S_{0}$ has the dimension of $\hbar$, the proportionality constants $\gamma_{i}$ should have the dimension of $\hbar^{-i}$. On the other hand in the case of a five--dimensional spacetime the five--dimensional gravitation constant $G^{(5)}$ in terms of units of length and units of $c$ and $\hbar$ reads
\begin{equation}
[G^{(5)}]=\frac{[c]^{3}L^{3}}{[\hbar]}.
\end{equation}

If we choose $G^{(5)}=c=k_{B}=1$, then the Planck constant in this units is of the order of $\hbar=(l_{p}^{(5)})^{3}$, where we have replaced $L$ with the five--dimensional Planck length $l_{p}^{(5)}$ \cite{zhu}. Note that this result is different to the four--dimensional case where $\hbar=(l_{p}^{(4)})^{2}$. It's evident now that the proportionality constants $\gamma_{i}$ should have the dimension of $r_{h}^{-3i}$ and multiplied by some other dimensionless constants $\beta_{i}$, as follows
\begin{equation}
\gamma_{i}=\beta_{i}r_{h}^{-3i}.
\end{equation}

Following Banerjee and Majhi and without going into details we assume the same form for the modified probability of the outgoing vector particles which can be written as follows
\begin{equation}
\tilde{\Gamma}=\exp\left[-\frac{4}{\hbar}\left(1+\sum_{i}\beta_{i}\frac{\hbar^{i}}{r_{h}^{3i}}\right)Im R\right].
\end{equation}
\bigskip

Thus for the quantum corrected Hawking temperature we find
\begin{equation}
\tilde{T}_{H}=\left(1+\sum_{i}\beta_{i}\frac{\hbar^{i}}{r_{h}^{3i}}\right)^{-1}T_{H}.\label{35}
\end{equation}

Note that the expansion (\ref{action}) to obtain corrections to all orders has been criticized \cite{yale}, where it was argued that there are no quantum corrections to the Hawking temperature via tunneling from a fixed background.  On the other hand, there were attempts to fix up the Banerjee and Majhi work and argued that this may have consequences to the information loss paradox \cite{singleton1,singleton2}. Next, we will try to calculate the quantum corrected entropy, so let us first write the first law of black hole mechanics for the charged black hole 
\begin{equation}
\tilde{S}=\int \frac{1}{\tilde{T}_{H}}\left(\mathrm{d}M-\Phi\,\mathrm{d}Q\right).\label{36}
\end{equation}

In order to calculate the quantum corrected entropy it is convenient to integrate with respect to $r_{h}$, instead of $M$ and $Q$ which is more difficult. To do so, let us consider the differential of the event horizon given by Eq. (\ref{10}), which gives 
\begin{equation}
2\left(\frac{r_{h}^{2}-Q^{2}}{r_{h}}\right)\mathrm{d}r_{h}=\mathrm{d}M-\Phi\,\mathrm{d}Q,\label{37}
\end{equation}
in which $M$ is the black hole mass and  $\Phi=-4Q \ln(r_{h})$ is the electric potential of the black hole near the horizon. On the other hand, the surfice area of the event horizon can be easley calculated as $A=\pi^{2}r_{h}^{3}$ \cite{chak2}, and hence the entropy of the black hole (in the units $G=c=1$, and by choosing the Boltzmann constant appropriately) takes the form $S=r_{h}^{3}/\hbar$. So $M$ can be obtained as a function of $S$ and $Q$ in the form \cite{chak2}
\begin{equation}
M=\left(S\,\hbar\right)^{2/3}-\frac{2}{3}Q^{2}\ln (S)+Const.
\end{equation}

The black hole temperature can be calculated from the energy conservation law of the black hole simply by differentiating $M$ with respect to $S$ at constant charge \cite{chak2}. It follows that 
\begin{equation}
T_{H}=\left(\frac{\partial M}{\partial S}\right)_{Q}=\frac{2 \hbar }{3}\left(\frac{1}{r_{h}}-\frac{Q^{2}}{r_{h}^{3}}\right).\label{39}
\end{equation}

If we insert Eqs. (\ref{35}),(\ref{37}) and (\ref{39}) into Eq. (\ref{36}) we end up with the following integral for the quantum corrected black hole entropy
\begin{equation}
\tilde{S}=\frac{3}{\hbar}\int \left(1+\sum_{i}\beta_{i}\frac{\hbar^{i}}{r_{h}^{3i}}\right)r_{h}^{2}\, \mathrm{d}r_{h}.
\end{equation}

Thus, by compering the last result with Eq. (\ref{36}) we see that the quantum corrected entropy can now be calculated easily by integrating with respect to $r_{h}$. Solving this integral we find
\begin{equation}
\tilde{S}=\frac{r_{h}^{3}}{\hbar}+3\beta_{1}\ln r_{h}-\frac{\hbar \beta_{2}}{r_{h}^{3}}+\dots
\end{equation}

If we make use of the Bekenstein--Hawking entropy $S_{BH}=S=r_{h}^{3}/\hbar$ \cite{chak2}, we can rewrite the last equation as
\begin{equation}
\tilde{S}=S_{BH}+\beta_{1}\ln S_{BH}-\frac{\beta_{2}}{S_{BH}}+Const+\dots
\end{equation}

Thus we have successfully recovered the logarithmic entropy correction term to the black hole entropy. Finally, we can now carry out the corrected specific heat capacity at constant charge, defined as
\begin{equation}
\tilde{C}=\tilde{T}_{H}\frac{\partial \tilde{S}}{\partial \tilde{T}_{H}}=\tilde{T}_{H}\frac{\partial \tilde{S}}{\partial r_{h}}\left(\frac{\partial \tilde{T}_{H}}{\partial r_{h}}\right)^{-1}.
\end{equation}

We find that 
\begin{equation}\label{51}
\tilde{C}=-\frac{3(r_{h}^{2}-Q^{2})(r_{h}^{6}+\hbar r_{h}^{3}\beta_{1}+\hbar^{2}\beta_{2})^{2}}{\hbar r_{h}^{3}(r_{h}^{8}-3Q^{2}r_{h}^{6}-2\hbar r_{h}^{5}\beta_{1}+\hbar^{2}(2Q^{2}-5 r_{h}^{2})\beta_{2})}+\dots
\end{equation}

In other words, the quantum corrected specific heat capacity goes to zero if
$r_{h}^{6}+\hbar r_{h}^{3}\beta_{1}+\hbar^{2}\beta_{2}=0$, at a certain $r_{h}$ which solves this equation. In Ref. \cite{banerjee1,banerjee2} it is argued that the numerical values of $\beta_{1}$ and $\beta_{2}$ are related to loop effects. Morover, using conformal field theory technique and by neglecting the $\hbar^2$ terms, i.e. $\beta_{2}=0$, it was shown that $\beta_{1}$ is given as follows \cite{banerjee1,banerjee2}
\begin{equation}
\beta_{1}=-\frac{1}{360 \pi}\left(-N_0-\frac{7}{4}N_{\frac{1}{2}}+13N_1+\frac{233}{4}N_{\frac{3}{2}}-212 N_{2}\right),
\end{equation}
where $N_{s}$ encodes the number of fields with spin `$s$'. If we assume this result is also valid in our case, we can take $N_1=1$ and $N_{\frac{1}{2}}=N_0=N_{\frac{3}{2}}=N_2=0$, then we find that quantum corrected specific heat capacity vanishes i.e. $\tilde{C}=0$ at $r_{h}=\left(\frac{13 \, \hbar}{360 \pi}\right)^{\frac{1}{3}}$. This seems to imply that the black hole cannot exchange radiation with the surrounding space at $r_{h}$, which may shed some light to the existence of a remnant. At this point, let us briefly mention that the existence of remnants is also supported by the GUP effects which are extensively used in the literature to calculate the quantum gravity corrections on the Hawking temperature, entropy, or specific heat capacity \cite{brito1,brito2,abdel1,abdel2,nasser1,nasser2,khalil1,khalil2,khalil3,khalil4}.

\section{Conclusion}

We have successfully calculated the Hawking temperature of uncharged massive bosons from 5D EYMGB black hole. We have solved the Proca equation by using the WKB approximation and the separation of variables and found five differential equations and constructed a $5\times 5$ matrix. We then extend our results by calculating the quantum correction to temperature, entropy, and specific heat capacity. We have started from the relation of the Planck constant and the five-dimensional Planck length in the units $G^{(5)}=c=1$ which are related by $\hbar=(l_{p}^{D=5})^{3}$. This on the other hand implies that the proportionality constants $\gamma_{i}$ are choosen to have the dimension of $r_{h}^{-3i}$. More specifically, we have recovered the first order logarithmic entropy correction beyond the semiclassical approximation by integrating with respect to $r_{h}$.  We first show that the Hawking temperature remains unaltered for this black hole configuration and the physical significance of this result is that black holes can also radiate massive vector particles and the mass/spin of the particles does not play any significant role in this process. On the other hand, the situation is a little bit different when we consider corrections beyond the semiclassical approximation. For instance, since $\beta_{1}$ and $\beta_{2}$ is shown to depend on the nature of spin of the particles, we have argued that the quantum corrected specific heat capacity goes to zero at some radius $r_{h}$ which may prevent a black hole from complete evaporation and hence the possibility of the existence of remnants. Even though the mass of the particles does no play any significant role, the spin of particles can be significant at the final stage of the black hole. We plan to extend our results in the near future in the spirit of Ref. \cite{xiang2} and expolore in more details the role of particles mass to the Hawking temperature under GUP effects.  
\bigskip
\section*{{Acknowledgements}}
The author is grateful to the editor and anonymous referees for their valuable and constructive suggestions.

\end{document}